\documentclass[aps,twocolumn,superscriptaddress,floatfix,preprintnumbers,nofootinbib]{revtex4-1}
\usepackage[utf8]{inputenc}  
\usepackage{amsmath}
\usepackage{amssymb} 
\usepackage{multirow}

\usepackage{enumerate}
\usepackage{amsmath}  
\usepackage{tikz}
\usepackage{tipa}
\usepackage{CJKutf8}
\usepackage{feynmf}
\usepackage{slashed}
\usepackage{braket}
\usepackage[toc,page]{appendix}
\usepackage{url}
\usepackage{natbib}
\usepackage{graphicx}
\usepackage[pdftex,bookmarks,linktocpage,pdfpagelabels,plainpages=false,hyperfigures,linkcolor=blue,citecolor=blue]{hyperref} 
\hypersetup{colorlinks=true}

\usepackage{mathrsfs,amssymb}  
\usepackage{cancel}
\usepackage[normalem]{ulem}
\usepackage{array}
\usepackage{booktabs}
\usepackage{verbatim}

\begin{document}
\begin{flushright}
MI-HET-824
\end{flushright}

\author{Doojin Kim}
\email{doojin.kim@usd.edu}
\affiliation{Department of Physics, University of South Dakota, Vermillion, SD 57069, USA}
\affiliation{Mitchell Institute for Fundamental Physics and Astronomy, Department of Physics and Astronomy, Texas A\&M University, College Station, TX 77843, USA}

\author{Jaehoon Yu}
\email{jaehoon@uta.edu}
\affiliation{Department of Physics, University of Texas, Arlington, TX 76019, USA}

\author{Jong-Chul Park}
\email{jcpark@cnu.ac.kr}
\affiliation{Department of Physics and Institute of Quantum Systems, Chungnam National University, Daejeon 34134, Republic of Korea}
\affiliation{Institute for Sciences of the Universe, Chungnam National University, Daejeon 34134, Republic of Korea}

\author{Hyunyong Kim}
\email{hyunyong.kim@cern.ch}
\affiliation{Department of Physics and Astronomy, Seoul National University, Seoul 08826, Republic of Korea}
\affiliation{Mitchell Institute for Fundamental Physics and Astronomy, Department of Physics and Astronomy, Texas A\&M University, College Station, TX 77843, USA}

\title{Beam-Dump Ceiling and Its Experimental Implication:\\
The Case of a Portable Experiment}

\begin{abstract} 
We generalize the nature of the so-called beam-dump ``ceiling'' beyond which the improvement on the sensitivity reach in the search for fast-decaying mediators dramatically slows down, and we point out its experimental implications that motivate tabletop-sized beam-dump experiments for the search.
Light (bosonic) mediators are well-motivated new-physics particles as they can appear in dark-sector portal scenarios and models to explain various laboratory-based anomalies. 
Due to their low mass and feebly interacting nature, beam-dump-type experiments, utilizing high-intensity particle beams can play a crucial role in probing the parameter space of such visibly decaying mediators, in particular, the ``prompt-decay'' region, where the mediators feature relatively large coupling and mass. 
We present a general and semianalytic proof that the ceiling effectively arises in the prompt-decay region of an experiment and show its insensitivity to data statistics, background estimates, and systematic uncertainties, considering a concrete example, the search for axion-like particles interacting with ordinary photons at three benchmark beam facilities: PIP-II at FNAL and SPS and LHC-dump at CERN. 
We then identify optimal criteria to perform a cost-effective and short-term experiment to reach the ceiling, demonstrating that very short-baseline compact experiments enable access to the parameter space unreachable thus far.
\end{abstract}

\maketitle

\section{Introduction} 
Although the Standard Model (SM) provides a successful description of elementary particles and their mutual interactions, many issues remain unresolved. 
For example, the nonvanishing masses of neutrinos and the existence of dark matter are unexplained by the SM, demanding new physics beyond the SM. 
Furthermore, various experimental anomalies, e.g., ATOMKI~\cite{Krasznahorkay:2015iga,Krasznahorkay:2019lyl,Krasznahorkay:2021joi,Krasznahorkay:2022pxs,Krasznahorkay:2023sax}, LSND~\cite{LSND:2001aii}, and MiniBooNE~\cite{MiniBooNE:2008yuf,MiniBooNE:2018esg,MiniBooNE:2020pnu}, 
together with recent experimental investigations by MicroBooNE relevant to the MiniBooNE excess (e.g., Refs.~\cite{MicroBooNE:2024tym,MicroBooNE:2025ntu}), may hint at the existence of new physics. 

Well-motivated explanations of these phenomena are deeply related to new particles that are very weakly or feebly interacting with the SM particles. 
In particular, new bosonic particles with MeV-range masses have been popular choices to address these observational anomalies and phenomena while satisfying the existing constraints. 
As a theoretical approach, the ideas of portal scenarios such as (pseudo)scalar and vector portals (see, e.g., Ref.~\cite{Batell:2022xau}) promote these MeV-scale particles to the mediators through which the dark-sector particles, including dark matter can communicate with the SM particles, and they often motivate MeV-scale light thermal dark matter.  
Therefore, searches for mediators of this sort can be regarded as an alternative avenue for dark-matter detection, especially in an accelerator-based experiment. 

Considering the expected mass range of these particles (henceforth collectively called mediators), fixed-target experiments such as beam-dump-type experiments are better suited than the energy-frontier facilities like the Large Hadron Collider (LHC) to probe MeV-scale physics as their beam energy is as large as $\mathcal{O}$(1-100 GeV).
Moreover, their high-instantaneous-intensity particle beams enable the production of feebly-interacting mediators.
For these reasons, the search prospects of (MeV-scale) Higgs-portal scalars, axion-like particles (ALPs), light $Z^\prime$, dark photons, etc., have been extensively investigated in the beam-dump-type experiments, especially, ongoing/upcoming/proposed experiments such as CCM, COHERENT, DUNE, FASER/FASER$\nu$, ICARUS, MicroBooNE, SBND, and SHiP (see also e.g., Refs.~\cite{Batell:2022xau,Ilten:2018crw,Bauer:2018onh,Fabbrichesi:2020wbt,Fortin:2021cog}, and references therein). 

A major goal of the experiments above is to expand the search as much as possible into the ``prompt-decay'' region, which none of the existing laboratory-based experiments or astrophysical/cosmological considerations have been able to constrain; a mediator belonging to this region would decay rather earlier due to its (relatively) large coupling and/or large mass value. 
A recent ambitious proposal DAMSA~\cite{Jang:2022tsp}---in which the detector system is placed extremely close to the signal source point (i.e., the beam target or dump\footnote{From now on, we will use {\it beam target} and {\it dump} interchangeably as we will show that the target width does not affect our results.})---showed the capability to start probing deeper into the prompt-decay region. 
Indeed, it has recently been demonstrated that for a given experimental configuration, there exists an inherent limitation dubbed the beam-dump ``ceiling'' beyond which the sensitivity improvement slows down dramatically so that an ordinary beam-dump-type experiment would have difficulty probing without a nonsensical increase of data statistics~\cite{Dutta:2023abe}. 
The actual location of the ceiling crucially depends on the beam energy and the detector distance from the beam target~\cite{Dutta:2023abe}. 

An important implication of the above observations with the beam-dump ceiling is that a collection of large amounts of data is no longer necessary to reach sensibly achievable sensitivity limits.
In this context, we delve into the robustness of the sensitivity reaches near the beam-dump ceiling against the overall/instantaneous beam intensity, the detector geometry (size and angular coverage), and underlying background/systematic uncertainty assumptions, taking three benchmark beam facilities, PIP-II~\cite{Ainsworth:2021ahm} at FNAL and SPS~\cite{Ahdida:2019ubf} and LHC-dump~\cite{Maestre:2021ajm} at CERN. 
From this investigation, we will demonstrate that a {\it tabletop-sized} detector system can probe the prompt-decay region of MeV-scale mediators rather {\it quickly}.
Therefore, one can imagine experimenting at one beam facility for a short period until the ceiling has been reached and taking the same detector system to a different beam facility to perform another experiment to expand the ceiling.
Inspired by DAMSA~\cite{Jang:2022tsp} now being proposed at PIP-II LINAC, we propose the concept of a portable DAMSA that could subsequently be performed at other facilities with various beam energies.

\begin{figure}[t]
    \centering
    \includegraphics[width=0.45\textwidth]{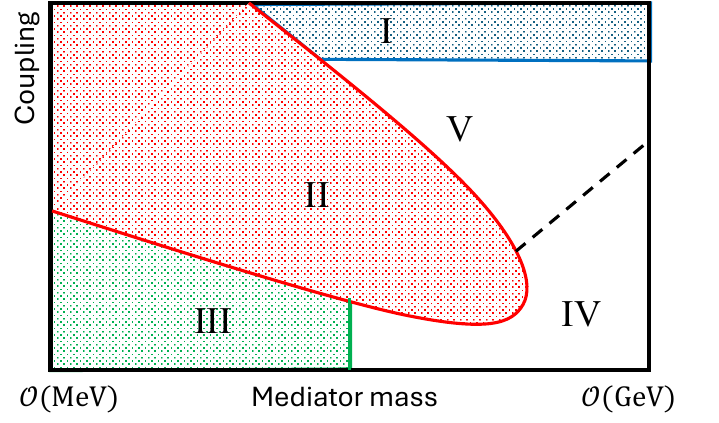}
    \caption{A schematic diagram of the existing constraints on generic mediators.}
    \label{fig:sensescheme}
\end{figure}

The rest of this paper is structured as follows. 
In Sec.~\ref{sec:ceiling}, we discuss the behavior of the beam-dump ceiling, examining in turn its dependence on the total beam intensity, backgrounds and systematic uncertainties, detector geometry, and instantaneous beam intensity. 
In Sec.~\ref{sec:study}, we validate the predicted ceiling behavior for the three benchmark beam facilities introduced above, focusing on ALPs coupled to the SM photon. 
We begin with a theory-level comparison in Sec.~\ref{sec:theorylevel}, adopting a common experimental configuration, including the beam-target and detector geometries. 
In Sec.~\ref{sec:practicallevel}, we present sensitivity projections for the SPS and LHC-dump configurations, considering feasible experimental setups---compatible with existing facility designs---and assuming a conservative level of reducible background suppression. 
Our conclusions appear in Sec.~\ref{sec:conclusions}. 

\section{Behavior of Beam-Dump ``Ceiling''} \label{sec:ceiling}

\subsection{Dependence on total beam intensity} 
We begin with a very rough sketch of the existing constraints of a generic (bosonic) mediator, say $\varphi$ (e.g., ALPs, dark photon, and light $Z'$), in FIG.~\ref{fig:sensescheme}.
Regions I and II are constrained by collider and beam-dump searches, respectively, whereas region III is constrained by various astrophysical and cosmological considerations including SN1987a, cosmic microwave backgrounds, and big-bang nucleosynthesis. 
Region IV is open and can be reached by increasing the statistics.
Our primary interest, however, is to extend the beam-dump reach to region V, i.e., the prompt-decay region mentioned earlier.

In generic beam-dump or fixed-target experiments, if a mediator is produced inside the dump or target, it should reach and decay into visible particles within the detector fiducial volume in order for it to be detected. 
This requirement can be formulated as a detection probability $P_{\rm det}$ in terms of the exponential decay law:
\begin{equation}
    P_{\rm det}=\exp\left(-\frac{L_{\rm dist}}{\tilde{l}_\varphi} \right) \left[1- \exp \left(-\frac{\Delta L_{\rm det}}{\tilde{l}_\varphi} \right) \right]\,,
\end{equation}
where $L_{\rm dist}$ and $\Delta L_{\rm det}$ denote the detector distance from the signal production point and the detector size along the direction of the mediator, respectively. 
$\tilde{l}_\varphi$ is the mean decay length of $\varphi$ in the laboratory frame. 
Note that the smaller the $L_{\rm dist}$, the larger the $P_{\rm det}$.
By definition, 
\begin{equation} 
L_{\rm dist},\Delta L_{\rm det} \gg \tilde{l}_\varphi \label{eq:condition}
\end{equation}
in the prompt-decay region, resulting in
\begin{equation}
P_{\rm det} \approx \exp\left(-\frac{L_{\rm dist}}{\tilde{l}_\varphi}\right).
\end{equation}

Suppose that for a given mass $m_\varphi$, an experiment can reach a coupling value $g$ with $N$ source particles inducing the $\varphi$ production. 
If data statistics increases (via beam intensity or beam exposure duration) by a factor of $X$, the new sensitivity reach $g^\prime$ is given by~\citep{Dutta:2023abe} 
\begin{equation}
    g^\prime \approx g \left(1+\frac{\log X}{\log(N\langle P_{\rm  prod} \rangle/N_{\rm req})-1}\right)^{1/2}\,, \label{eq:ceiling} 
\end{equation}
where $N_{\rm req}$ is the number of signal events required to define the sensitivity reach $g$, and $\langle P_{\rm prod} \rangle$ is the average mediator production probability at $g$ for the given source particle. 
For a large background, $\log X$ can be replaced with $\log X^{1/2}$. 

In ongoing/planned beam-dump or fixed-target experiments, particle beams are highly intensified, hence $N$ is very large. 
In the prompt-decay region, the coupling is not too small and so is $\langle P_{\rm prod} \rangle$.
Finally, unless an experiment suffers from poor background rejection or poor understanding of systematic uncertainties, $N_{\rm req}$ need not be too large. 
Therefore, the denominator in Eq.~\eqref{eq:ceiling} is larger than the numerator, as long as $X$ is at a sensible level. 
This implies that any feasible level of an increase in data collection would not allow for an appreciable improvement in the sensitivity reach. 
This conversely implies that {\it decreasing data collection would not result in a considerable degradation in the sensitivity reach}, as long as $|\log X| \ll \log(N\langle P_{\rm prod} \rangle/N_{\rm req})-1$. 
The overall decrease in data collection can be accomplished in two ways: 
\begin{itemize} 
    \item[($a$)] smaller beam exposure, and
    \item[($b$)] smaller beam power.
\end{itemize}

\subsection{Dependence on background and systematic uncertainty estimates}  
A similar exercise can be done for the case where a different number of signal events $N_{\rm req}^{\prime\prime}$ is required to determine the sensitivity reach. 
For example, one could incorrectly estimate backgrounds and/or systematics so that the true sensitivity reach would have been $g^{\prime\prime}$. 
We find that $g^{\prime\prime}$ is expressed as
\begin{equation}
    g^{\prime\prime} \approx g \left(1+ \frac{\log(N_{\rm req}^{\prime\prime}/N_{\rm req})}{\log (N \langle P_{\rm prod} \rangle/N_{\rm req})-1} \right)^{-1/2}\,. \label{eq:cond2}
\end{equation}
This relation suggests that unless an estimate of backgrounds or systematics is very poorly done, the numerator $|\log(N_{\rm req}^{\prime\prime}/N_{\rm req})|$ is much smaller than the denominator $ \log(N\langle P_{\rm prod} \rangle/N_{\rm req})-1$, making $g^{\prime\prime}$ differ insignificantly from $g$. 

All these observations demonstrate that {\it the sensitivity near the beam-dump ceiling is robust and nearly insensitive to details of backgrounds, systematics, and beam intensity for a given experimental configuration} unless they are grossly misestimated. 
We will verify this expectation shortly in a (semi)quantitative way, taking the ALP interacting with the SM photon as a concrete example. 

\subsection{Dependence on detector geometry}
Many experiments that aim to detect mediator decay signals often accompany a finite-size decay volume or equivalent module immediately upstream of the particle detection modules (e.g., DAMSA~\cite{Jang:2022tsp}, DarkQuest~\cite{Apyan:2022tsd}, NA62~\cite{NA62:2017rwk}, SHiP~\cite{Alekhin:2015byh}, etc). 
Obviously, the mediators residing near the ceiling are likely to decay promptly, and thus a long decay volume is unessential to probe this region. 
More specifically, decay events contributing to the sensitivity determination near the ceiling mostly happen at the beginning of the decay volume. 
Therefore, the sensitivity reaches in the prompt-decay region are nearly unaffected by the length of the decay volume. 

Furthermore, the prompt mediators that reach and decay in the decay volume are likely to be highly boosted.
Mediators with a large boost factor are induced by high-energetic, more forward-directed source particles. 
Therefore, wide angular coverage is less motivated, in order for a detector to explore the prompt-decay region. 

Considering the above two geometry-wise observations, which can be summarized as 
\begin{itemize}  
    \item[($c$)] shorter length of the decay volume, and
    \item[($d$)] smaller angular coverage of the detector system, 
\end{itemize}
we expect that a tabletop-sized detector system can, in principle, be sufficient to probe the prompt-decay region and reach its own beam-dump ceiling. 
We will show the dependence of sensitivity reaches on these aspects again in the context of the ALP scenario. 

\subsection{Dependence on instantaneous beam intensity}
The discussion thus far is relevant to any generic mediators, irrespective of whether their visible decay products involve some features that are distinguishable from potential (reducible) backgrounds.  
Depending on the final-state features in the mediator decay, an appropriate choice of instantaneous beam intensity (i.e., single-bunch intensity) would further reduce the required beam exposure to achieve the aimed sensitivity goal in the prompt-decay region.
For example, in the DAMSA proposal~\cite{Jang:2022tsp}, the authors studied the ALP decaying to a photon pair and identified an accidental overlap of a pair of beam-related-neutron-induced photons that mimics the diphoton signal as the primary background. 

One can expect that the number of background events will be reduced if the number of beam-related-neutron-induced photons in a single beam bunch length decreases. 
For a more quantitative calculation, suppose that a given protons-on-target (PoT) is delivered by $n_{\rm bunch}$ beam bunches, and each bunch results in $n_\gamma$ neutron-induced photons (possibly after a set of cuts). 
Since the number of photon pairs within the given beam bunch is $\sim n_\gamma^2$, the total number of signal-faking accidental photon pairs in the given PoT is $\sim n_{\rm bunch} n_\gamma^2$. 
Therefore, if the same PoT is delivered by $f$ times more bunches, the number of beam-related-neutron-induced photons in a beam bunch is $\sim n_\gamma/f$, and the expected total number of backgrounds will be reduced by a factor of $f$ to $\sim n_{\rm bunch} n_\gamma^2/f$.

This simple algebra clearly shows that while backgrounds such as the beam-related-neutron-induced one could be formidable, an experiment to perform a search for featureful mediator signals could benefit greatly in reaching its ceiling at a lower beam exposure with
\begin{itemize}  \itemsep1pt \parskip0pt \parsep0pt
    \item[($e$)] smaller instantaneous beam intensity.
\end{itemize}
As this is related to backgrounds, we study the sensitivity dependence on the instantaneous beam intensity by varying background estimates.

\section{Benchmark study} \label{sec:study}
We are now in the position to validate the arguments thus far in the context of actual analyses for an example physics case. 
To this end, we consider ALPs interacting with the SM photon:
\begin{equation}
    -\mathcal{L}_{\rm int} \supset \frac{1}{4} g_{a\gamma\gamma} a F_{\mu\nu}\tilde{F}^{\mu\nu}\,,
\end{equation}
where $a$, $F_{\mu\nu}$ ($\tilde{F}_{\mu\nu}$), and $g_{a\gamma\gamma}$ are the ALP field, the (dual-)field strength tensor of the SM photon, and the interaction strength between the ALP and the SM photon, respectively.  
Under this Lagrangian, an ALP is produced by the Primakoff scattering process of the photons~\cite{Primakoff:1951iae} inside the beam target, and it decays into a photon pair inside the experimental system. 

\subsection{Theoretical benchmark comparison of the three facilities} \label{sec:theorylevel}

\begin{table}[t]
    \centering
    \resizebox{\columnwidth}{!}{
    \begin{tabular}{c|c c c| c}
    \hline \hline
      Beam & $E_p$ [GeV] & PoT/bunch & PoT/yr & $E_{\gamma,{\rm  cut}}$ [GeV]\\
      \hline
      PIP-II~\cite{Ainsworth:2021ahm} & 0.8 & $7.7 \times 10^7$  & $4\times 10^{23}$ & 0.1\\
      SPS~\cite{Ahdida:2019ubf,Albanese:2878604} & 400 & $1.7\times 10^{5}$ & $4\times 10^{19}$ & 5 \\
      LHC-dump~\cite{Maestre:2021ajm} & 7,000 & $2.2\times 10^{11}$ & $1.1\times 10^{17}$ & 20 \\
      \hline \hline
    \end{tabular}
    }
    \caption{Benchmark beam facilities: Key beam parameters (second through fourth columns) and our analysis cuts on photons at ECAL (fifth column).}
    \label{tab:beams}
\end{table}

The three benchmark beam facilities and their key specifications are summarized in Table~\ref{tab:beams}. 
The default bunch rate (henceforth denoted by $R_b^\star$) can be calculated from PoT/bunch in this table. 
For PIP-II, a range of bunch repetition rates is possible~\cite{Ainsworth:2021ahm}, but we assume 100 MHz for illustration. 
SPS features a slow beam spill of $4\times 10^{13}$ protons over 1.2 seconds. 
We assume that the beam is extracted in bunches evenly spaced at 200~MHz~\cite{SHiP:2021nfo}, corresponding to $1.7\times 10^5$ PoT per bunch. 
By contrast, the proton beams are dumped to the LHC-dump 2,756 times per day, with $2.2\times 10^{11}$ PoT per bunch~\cite{Maestre:2021ajm}. 
This corresponds to a yearly exposure of $1.1 \times 10^{17}$ PoT assuming a 50\% duty factor. 

The photon flux expected in each of the beam facilities is estimated using \texttt{GEANT4}~\cite{GEANT4:2002zbu}. 
We assume a tungsten dump for the three beam facilities for comparison purposes, but our conclusions do not depend on the choice of the dump as long as the material is the same. 
The selected target lengths are 1~meter (PIP-II), 1.2~meters (SPS), and 1.2~meters (LHC-dump), and our simulation shows that the mean photon production positions are at $\sim 5$~cm, $\sim 15$~cm, and $\sim 20$~cm after $E_{\gamma,{\rm cut}}$ in Table~\ref{tab:beams}, respectively. 
We then envision a similar experimental setup as in the DAMSA proposal~\cite{Jang:2022tsp}: a certain length of vacuum decay chamber follows from the beam-target system and an electromagnetic calorimeter (ECAL) is placed downstream of the decay chamber.
We further assume a 100 MeV photon energy threshold and a 1$^\circ$ angular resolution of the detector although the detailed ECAL capabilities hardly affect our conclusions. 
Finally, we follow the same data analysis scheme and beam-related-neutron-induced background suppression strategy as explicated in Ref.~\cite{Jang:2022tsp}. 

\begin{figure*}[t]
    \centering
    \includegraphics[width=0.45\textwidth]{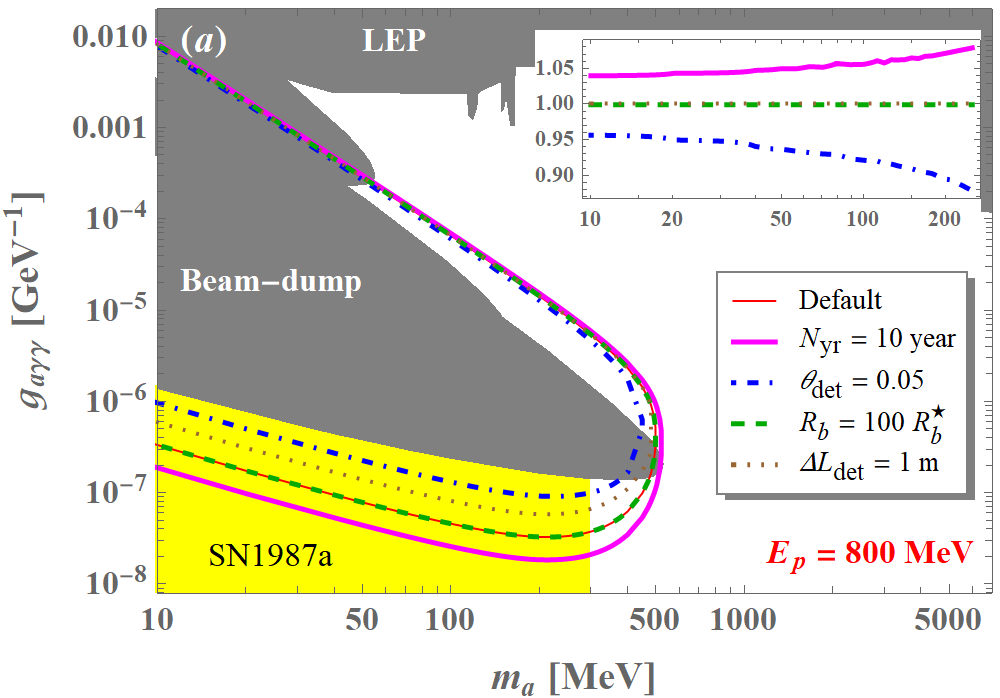} \hspace{0.1cm}
    \includegraphics[width=0.45\textwidth]{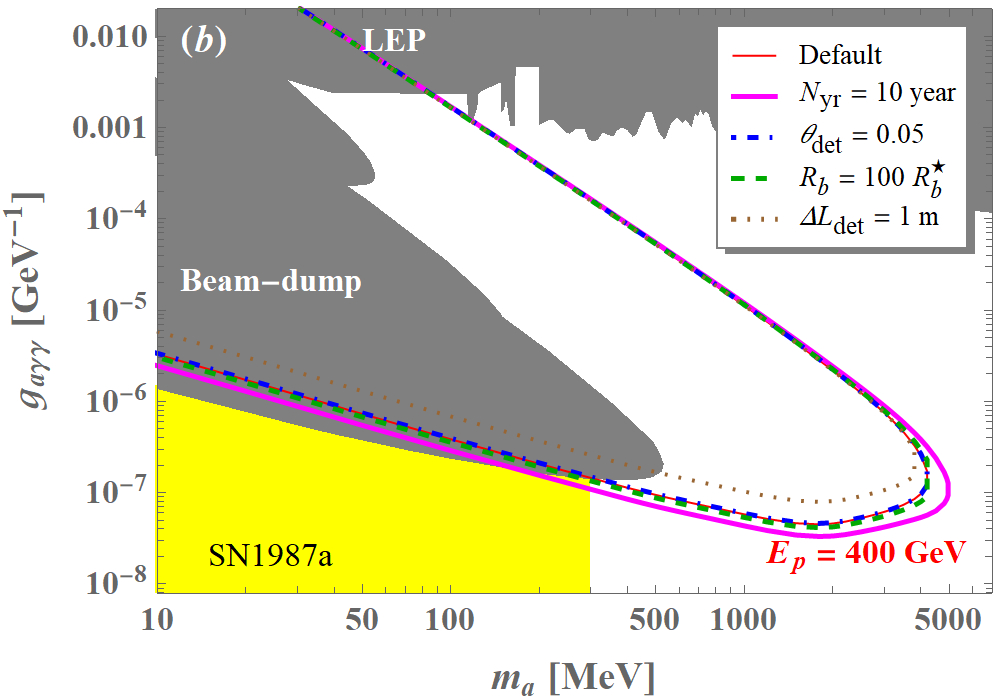} \\
    \includegraphics[width=0.45\textwidth]{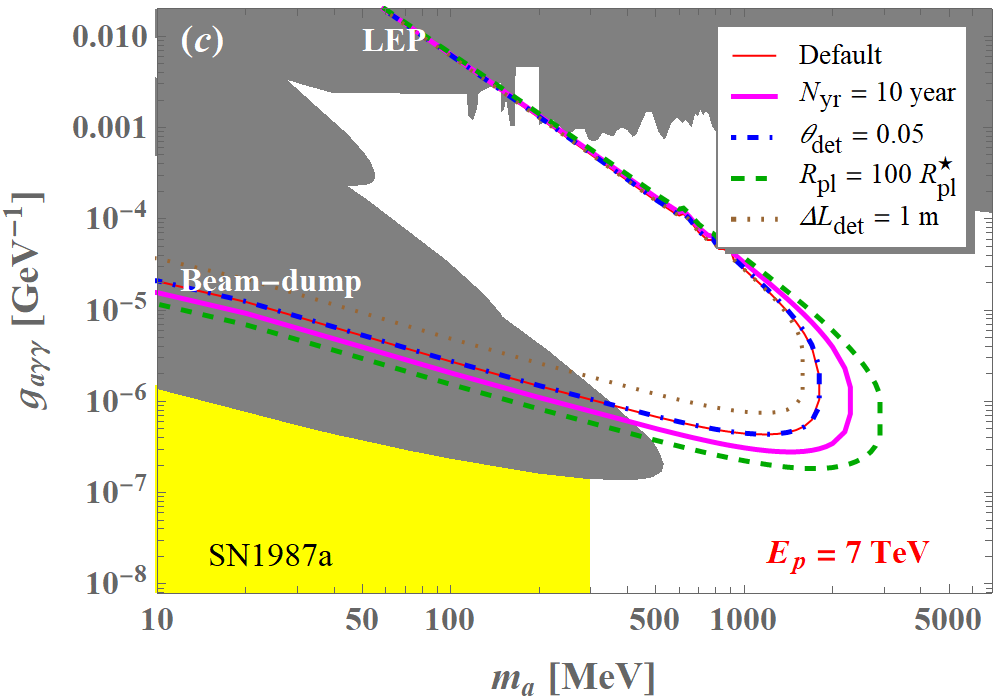} \hspace{0.1cm}
    \includegraphics[width=0.45\textwidth]{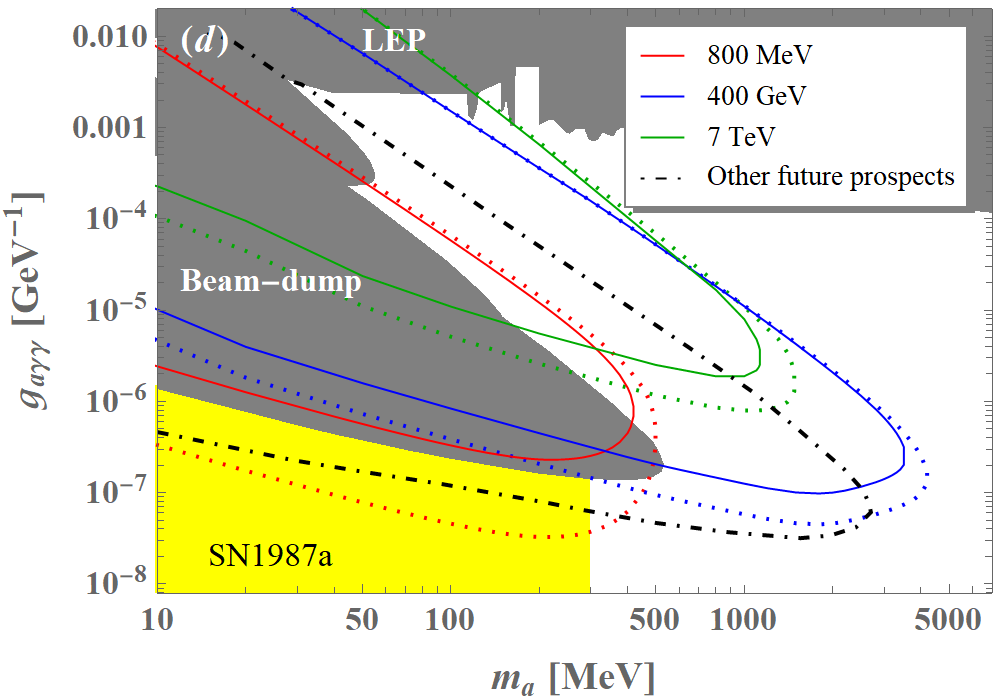}
    \caption{Expected 90\% C.L. sensitivity reaches for coupling $g_{a\gamma\gamma}$ as a function of ALP mass $m_a$ at ($a$) PIP-II, ($b$) SPS, and ($c$) LHC-dump. 
    The inset in panel ($a$) shows the fractional change of the sensitivity with respect to the default case. 
    We compare the three benchmark beam facilities for $\Delta L_{\rm det}=1$ meter, $\theta_{\rm det}=0.05$~rad, and $N_{\rm yr}=3$~months in panel ($d$) where the sensitivity reaches with the default setup are shown by dotted lines [i.e., the red solid curves in panels ($a$), ($b$), and ($c$)] for reference purposes. 
    In panel ($d$), the black dot-dashed line represents other future prospects, including DarkQuest~\cite{Apyan:2022tsd}, DUNE~\cite{Brdar:2020dpr}, FASER/FASER2~\cite{Feng:2018pew}, LDMX~\cite{Berlin:2018bsc}, and SHiP~\cite{Albanese:2878604}, for comparison.}
    \label{fig:sense}
\end{figure*}

We first consider PIP-II providing an 800-MeV proton beam, and we show the expected 90\% C.L. sensitivity reaches under various experimental environments in FIG.~\ref{fig:sense}($a$). 
The reference or default parameters are defined as follows:
\begin{itemize}  
    \item detector distance from the mean photon production position in the target, $L_{\rm dist}=1$~meter,
    \item length of the decay volume $\Delta L_{\rm det}=10$~meters,
    \item half the detector opening angle $\theta_{\rm det}=0.5$~rad,
    \item beam bunch rate $R_b$ to deliver the given PoT,
    \item beam exposure $N_{\rm yr}=1$~year.  
\end{itemize}
In addition to these parameter choices, we impose an energy cut on signal photons, i.e., $E_{\gamma, {\rm cut}}=100$~MeV, to mitigate the neutron-induced photon backgrounds and to reflect the reasonable detector capabilities. We find that a negligible number of accidental photon pairs can pass this hard energy cut. We do not vary the detector distance but fix it to be $L_{\rm dist}=1$~meter throughout our study since we are primarily interested in the prompt-decay region.\footnote{Strictly speaking, signal production points differ from event to event, hence $L_{\rm dist}$ varies. We consider this as an average distance.} 
The thin red solid curve shows our reference sensitivity reaches expected at the 800 MeV PIP-II. 
For comparison purposes, the current constraints compiled in e.g., Refs.~\cite{Fortin:2021cog,Batell:2022xau} are in the grey and yellow shaded area; they include Belle-II~\cite{Belle-II:2020jti}, CHARM~\cite{CHARM:1985anb}, E137~\cite{Bjorken:1988as}, E141~\cite{Riordan:1987aw}, LEP~\cite{OPAL:2002vhf}, LHC (Pb)~\cite{CMS:2018erd,ATLAS:2020hii}, NA64~\cite{NA64:2020qwq}, NuCal~\cite{Blumlein:1990ay}, and PrimEx~\cite{Aloni:2019ruo}.

The remaining four curves compare the sensitivity reaches with different experimental parameter choices against the reference one, keeping all others at default values. 
The magenta solid line is for a 10-times-increased beam exposure. 
It enables extending the sensitivity reaches toward the bottom and the higher-mass regions (region IV in FIG.~\ref{fig:sensescheme}) that are sensitive to statistics. 
The sensitivity toward the prompt-decay region (region V in FIG.~\ref{fig:sensescheme}), however, does not change since the limit has already been reached. 
A similar statement is relevant to the other cases: a 10-times-decreased polar-angular coverage with the blue dot-dashed line, a 100-times-increased bunch rate with the green dashed line, and a 10-times-decreased decay volume length with the orange dotted line.
Again for all of them, we lose or gain some sensitivity in the floor and higher-mass regions (region IV), but their impacts on the sensitivity in the prompt-decay region (region V) are negligible.

All these comparisons suggest that sensitivity estimates near the beam-dump ceiling (below $m_a \lesssim 300$~MeV for PIP-II) are robust as predicted earlier, changing at most within the $5- 10\%$ level below $m_a = 300$~MeV, as shown in the inset.
More quantitatively speaking, for example, the comparison between the red and green curves essentially represents the case where the required number of signal events to determine the sensitivity differs by an order of magnitude (i.e., a 1,000\% change) due to a mismodeling of systematics and/or an incorrect background estimation. 

Figures~\ref{fig:sense}($b$) and \ref{fig:sense}($c$) display our sensitivity estimates at SPS and LHC-dump, respectively, assuming a fixed baseline of $L_{\rm dist}=1$~m. 
While facility-specific studies are necessary to determine the optimal baselines in these cases, we adopt this value for illustrative purposes to enable a uniform comparison across different beam facilities.
Unlike PIP-II, more energetic photons would come out of the target in addition to the neutron-induced photons, such as those from processes like $K_L^0\to3\pi^0\to6\gamma$, due to higher proton beam energies. 
Our \texttt{GEANT4} simulation suggests that the backgrounds from photons exiting the target are significantly suppressed with $E_{\gamma,{\rm cut}}$ and PoT/bunch in Table~\ref{tab:beams}; more quantitatively, there are $\sim 1.2\times 10^9$ and $\sim 4.2\times 10^{12}$ accidental photon pairs per year at SPS and LHC-dump, respectively. 
Assuming a similar-level suppression factor of $\sim 10^8-10^9$ as in Ref.~\cite{Jang:2022tsp} along with posterior cuts (e.g., invariant mass window) therein, we expect $\sim 12$ and $\sim 42,000$ diphoton backgrounds annually. The effectiveness of the posterior cuts such as the distance-of-closest-approach cut, arrival-time-difference cut, fiducial-volume cut, trace-back cut, and invariant-mass-window cut, as developed in Ref.~\cite{Jang:2022tsp}, relies on the detector's capabilities. 
For illustration, we assume a similar level of detector performance to achieve comparable background suppression.

Similarly to the PIP-II case, the maximum sensitivity reaches around the beam-dump ceiling are rather robust and nearly insensitive to beam parameters, detector geometry, statistics, etc.
As pointed out earlier, the impact of the detector angular coverage is much less pronounced in SPS and the LHC than that in PIP-II since the signal is more highly forward-directed.

Along these lines, our study further enables the identification of key experimental requirements for exploring the ``prompt-decay'' parameter space, suggesting that a compact ``{\it tabletop-sized}'' detector is sufficient to reach the beam-dump ceiling of a given experiment with a {\it moderate beam exposure}. 
Therefore, multiple, short-term experiments with the same compact detector system can be performed at different beam facilities, providing complementary information for probing the parameter space, especially toward the prompt-decay regime. 
For example, the solid curves in FIG.~\ref{fig:sense}($d$) show the sensitivity of the three beam facilities with $\Delta L_{\rm det}=1$~meter, $\theta_{\rm det}=0.05$~radians, and $N_{\rm yr}=3$~months. We employ an ECAL with just a 10 cm-scale radius for these sensitivity estimates. For comparison, we also show the sensitivity curves corresponding to the default configurations by dotted lines. 
These examples clearly support the case of compact, portable experiments that can be performed in different facilities, complementing each other in expanding reachable parameter space, possibly beyond other future prospects, including DarkQuest~\cite{Apyan:2022tsd}, DUNE~\cite{Brdar:2020dpr}, FASER/FASER2~\cite{Feng:2018pew}, LDMX~\cite{Berlin:2018bsc}, and SHiP~\cite{Albanese:2878604}, represented by the black dot-dashed line.

We stress that Fig.~\ref{fig:sense}($d$) is not a facility-optimized projection, but a controlled benchmark in which key experimental inputs are held fixed to isolate the impact of beam energy on production kinematics and acceptance. 
Because the achievable (instantaneous) beam exposure, beam-time structure, and background environment are highly facility-dependent, the relative ordering of reaches can shift once these effects are included. 
Therefore, Fig.~\ref{fig:sense}($d$) should be read as demonstrating complementarity in kinematics/acceptance rather than as evidence for a universal optimal beam energy.

\subsection{Practical comparison of the SPS and LHC-dump facilities} \label{sec:practicallevel}

As mentioned earlier, the study in the previous section is largely theoretical for the SPS and LHC-dump facilities and will require more detailed investigations to assess the ECAL’s exposure to unwanted particles, including escaping hadrons (e.g., beam-related neutrons) and muons. 
While we leave a full feasibility study of this type for future publications, including the DAMSA Pathfinder work~\cite{DAMSA:2026fiw}, we present here sensitivity estimates based on more practical configurations that could plausibly be achieved at the two facilities:
\begin{itemize}
    \item {\bf SPS}: We leverage the planned SHiP experimental setup~\cite{Albanese:2878604} to test the DAMSA proposal at the SPS. 
    SHiP employs a $\sim 1.4$-meter TZM (titanium-zirconium-doped molybdenum) $\oplus$ tungsten target, followed by a $\sim40$-meter scale magnetic field area to bend away the muons escaping from the target. 
    The collaboration’s dedicated muon-shielding simulation indicates the existence of a nearly muon-free ``sweet spot'' located about 15 m downstream from the front of the shielding region~\cite{SHiPTalk, private}. 
    Given the mean photon production point, the effective baseline is $\sim 16$ meters.

    \hspace{0.5cm} A full simulation incorporating both the beam target and the entire shielding region is computationally prohibitive. 
    Instead, we simulate the target followed by an iron yoke with four representative lengths---0.25~m, 0.5~m, 1~m, and 2~m---using $10^6$ incident protons in each configuration. 
    We find that the number of photons exiting the yoke with $E_\gamma > 5$~GeV decreases rapidly with increasing yoke length, and the trend is well described by a straight line in log space. 
    Extrapolating to a 15~m iron yoke yields an expected photon rate below $10^{-22}$ per incident proton. 
    Given the POT per bunch and the annual beam intensity, this corresponds to $\sim 10^{-8}$ photons per bunch, implying $\sim 0.01$ accidental photon pairs per year. 
    Therefore, under this setup, a nearly background-free analysis is feasible without imposing additional {\it a posteriori} cuts.

    \item {\bf LHC-dump}~\cite{Maestre:2021ajm}: Once the luminosity decays below an operational limit, the fill is intentionally terminated, and the circulating protons are extracted to the $\sim 8$ meter-scale graphite beam-dump block, followed by a $\sim 3$ meter thick concrete to prepare for the next injection cycle. 
    A few-meter-long empty space is located downstream of the concrete module. 
    We envision placing an additional 1.5 m tungsten dump in this region, followed by the DAMSA detector. 
    Our \texttt{GEANT4} simulation indicates that most dumped-proton interactions occur at $\sim 2.5$ meters, so the effective baseline is 10 meters under this setup.

    \hspace{0.5cm} Our simulation further suggests that, per proton bunch containing $2.2\times 10^{11}$ protons, 17 photons with energies above 20 GeV enter the DAMSA detector region. 
    This results in 136 accidental photon pairs, some of which could potentially mimic the ALP signal events, yielding approximately $6.8\times 10^7$ diphoton-like events annually. 
\end{itemize}

We are now ready to present the sensitivity estimates for the SPS and LHC-dump configurations, assuming the experimental setup described above. 
For comparison, we adopt the same detector size as in the $\theta_{\rm det}=0.05$ rad case: a calorimeter region with a length of order 1~m and a radius of order 10~cm. 
With this geometry, the DAMSA detector subtends angles of 5.9~mrad and 9.1~mrad with respect to the beam axis for the SPS and LHC-dump cases, respectively.

\begin{figure}[t]
    \centering
    \includegraphics[width=0.9\linewidth]{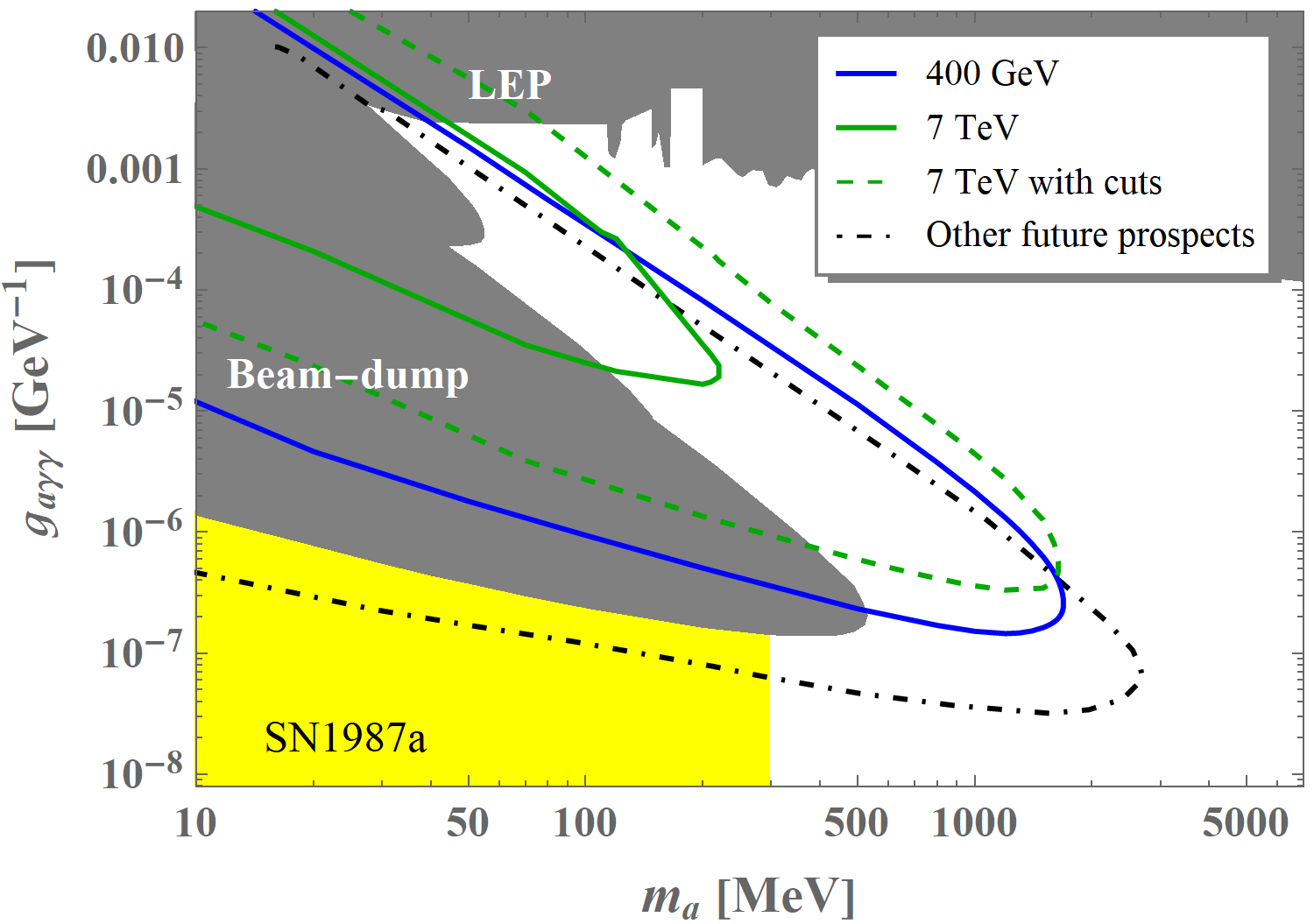}
    \caption{Expected 90\% C.L. sensitivity reaches for coupling $g_{a\gamma\gamma}$ as a function of ALP mass $m_a$ at SPS (blue curve) and LHC-dump (green curves), assuming the experimental configuration described in the text. 
    To illustrate the maximal potential of the LHC-dump configuration, we also show a sensitivity estimate (green dashed curve) assuming a nearly background-free environment. 
    The black dot-dashed line represents other future prospects as in FIG.~\ref{fig:sense}. }     \label{fig:realsensitivity}
\end{figure}

The blue solid curve in FIG.~\ref{fig:realsensitivity} shows our SPS sensitivity estimates assuming a one-year exposure, while the green solid curve corresponds to the LHC-dump case with no cuts applied other than $E_\gamma>20$~GeV. 
The black dot-dashed line represents other future prospects as before. 
Since the LHC-dump case (green solid curve) is overwhelmed by a formidable amount of reducible backgrounds, its full potential is not adequately reflected. Indeed, we observe that, in the experimental configuration under consideration, the LHC-dump case would not reach its own ``ceiling,'' since the beam intensity does not appear to be sufficient for the experiment to become limited by the beam-dump ceiling rather than by backgrounds. 
In other words, reducing the backgrounds would allow one to probe deeper into the prompt-decay region, potentially up to the ceiling. We therefore also show the maximal reach, indicated by the green dashed curve, which could be achieved with an appropriate set of cuts. 
For example, the set of kinematic cuts proposed in Ref.~\cite{Jang:2022tsp} can suppress accidental diphoton-like backgrounds by roughly $8-9$ orders of magnitude, yielding a nearly background-free environment. 
However, validating this expectation requires a careful, dedicated analysis, which we defer to future work as mentioned earlier.

\section{Conclusions} \label{sec:conclusions}

In this paper, we presented a study of the (visibly) decaying signals of new mediators at beam-dump-type experiments, taking ALPs with couplings to photons for illustration. 
We clearly showed that once a given experiment reaches its inherent ceiling, the sensitivity near the beam-dump ceiling, the maximum sensitivity reach that an experiment can accomplish in the prompt-decay region, is robust and insensitive to detector geometry, beam specifications, and background/systematics estimates, regardless of the detector baseline. 
Moreover, we provided means to assess whether the projected sensitivity of an experiment hits the ceiling in Eqs.~(\ref{eq:condition})--(\ref{eq:cond2}).

We also considered more realistic background levels and experimental configurations compatible with existing SPS and LHC-dump facility designs. 
Our sensitivity estimates still indicate that a tabletop-sized DAMSA detector can probe new regions of ALP parameter space, potentially extending beyond other future prospects. 
These results clearly motivate further dedicated studies, including the exploration of additional physics cases at these facilities.

We hope that DAMSA can be realized in the near future~\cite{DAMSA:2026fiw}, probing unexplored and challenging regions of dark-sector parameter space and potentially uncovering hints of new physics.

\medskip

\section*{Acknowledgments} 
For facilitating portions of this research, the authors wish to acknowledge the Center for Theoretical Underground Physics and Related Areas (CETUP*), the Institute for Underground Science at the Sanford Underground Research Facility (SURF), and the South Dakota Science and Technology Authority for hospitality and financial support, as well as for providing a stimulating environment.
This article was supported by the computing resources of the Global Science Experimental Data Hub Center (GSDC) at the Korea Institute of Science and Technology Information (KISTI).
We thank Wooyoung Jang of the University of Texas at Arlington for valuable discussions.
The work of DK is supported in part by the U.S.~Department of Energy Grant DE-SC0010813. 
The work of JCP is supported by the National Research Foundation of Korea (NRF) grant funded by the Ministry of Science and ICT [RS-2024-00356960] and by the Ministry of Education [RS-2025-25442707].
The work of JY is supported by the University of Texas at Arlington, U.S. Department of Energy under Grant No. DE-SC0011686 and NRF grant [RS-2024-00350406]. 
The work of HK is supported by the U.S.~Department of Energy under Grant No.~DE-SC0010813 and NRF grant [RS-2026-25493297 and RS-2024-00350406].

\bibliography{ref}

\end{document}